\documentclass[prb,reprint,showpacs,amsmath,amssymb,superscriptaddress,citeautoscript,longbibliography]{revtex4-1}

\usepackage[plainpages=false,pdfpagelabels,colorlinks=true,linkcolor=red,urlcolor=blue,citecolor=blue,pdftitle={Title},pdfauthor={},pdfdisplaydoctitle=true,pdfduplex=DuplexFlipLongEdge]{hyperref}
\usepackage{epsfig}
\usepackage{amsmath,amssymb}
\usepackage{physics}
\usepackage{graphicx}
\usepackage{subcaption}
\usepackage[dvipsnames,usenames]{color}
\usepackage[normalem]{ulem}
\usepackage{soul}               
\usepackage{caption}
\usepackage{ragged2e}

\begin{document}

\title{Learning by Confusion:  The Phase Diagram of the Holstein Model}

\author{George Issa}
\affiliation{Department of Physics and Astronomy, University of California, Davis,
  California 95616, USA}
\author{Owen Bradley}
\affiliation{Department of Physics and Astronomy, University of California, Davis,
  California 95616, USA}
\author{Ehsan Khatami}
\affiliation{Department of Physics and Astronomy, San Jos\'{e} State University, San Jos\'{e}, CA 95192 USA}
\author{Richard Scalettar}
\affiliation{Department of Physics and Astronomy, University of California, Davis,
  California 95616, USA}

\begin{abstract}
We employ the “learning by confusion” technique, an unsupervised machine learning approach for detecting phase transitions, 
to analyze quantum Monte Carlo simulations of the two-dimensional Holstein model—a fundamental model for electron-phonon interactions on a lattice. 
Utilizing a convolutional neural network, we conduct a series of binary classification tasks to identify Holstein critical points based on the neural network's learning accuracy. 
We further evaluate the effectiveness of various training datasets, including snapshots of phonon fields and other measurements resolved in imaginary time, for predicting distinct 
phase transitions and crossovers. Our results culminate in the construction of the finite-temperature phase diagram of the Holstein model. 
\end{abstract}

\pacs{
71.30.+h,   
71.45.Lr, 	
63.20.-e    
}

\maketitle

\section{Introduction}

Artificial Intelligence (AI) and Machine Learning (ML) approaches have emerged as a 
powerful technique to study classical and quantum phase transitions
(often using the output of Monte Carlo simulations as training data)
~\cite{wang2016discovering,carrasquilla2017machine,hu2017discovering,ch2017machine,carleo2017solving,zhang2017quantum,ch2018unsupervised,wetzel2017unsupervised,huembeli2018identifying}, 
out-of-equilibrium phenomena~\cite{venderley2018machine,bohrdt2021analyzing,zhang2022anomalous,rodriguez2022quantum}, 
and also including the use of experimental data\cite{bohrdt2019classifying,rem2019identifying,zhang2019machine,torlai2019integrating,khatami2020visualizing,samarakoon2020machine}.
We refer the interested reader to recent reviews that provide comprehensive overviews of applications of AI/ML to strongly-correlated models\cite{johnston2022perspective,dawid2022modern,carrasquilla2020machine}.

In the context of exploring itinerant electron Hamiltonians, one
focus of ML approaches has been on the Hubbard model and understanding
magnetic, charge, and exotic ($d$-wave) pairing correlations as well as pseudogap and strange metal 
phases\cite{bohrdt2019classifying,nomura2017restricted,broecker2017quantum,shinjo2019machine,canabarro2019unveiling,khatami2020visualizing,striegel2023machine,xiao2024extracting},
whereas ML investigations of electron-phonon Hamiltonians are somewhat more limited\cite{chen18,li2019accelerating,nomura2020machine,cheng2023machine}.
The Holstein model\cite{holstein1959} has a
phenomenology characterized by
charge density wave (CDW) order at commensurate filling on a bipartite lattice.  This insulating phase gives way to
conventional ($s$-wave) pairing upon doping.
These phases have been extensively studied with quantum simulations and conventional
methods of analysis, i.e.~the evaluation of order parameters and their finite size scaling
\cite{peierls79,hirsch82,hirsch83,scalettar89,marsiglio90,freericks93,ohgoe17,hohenadler19,bradley21,nosarzewski21,araujo22}.
Several subtle effects emerge, including a non-monotonic dependence of the superconducting
transition temperature on the electron-phonon coupling strength
\cite{weber2018,zhang2019,feng2020,chen2019,bradley21},   
a behavior at variance with Eliashberg theory\cite{esterlis18}. Unlike the Hubbard model, the
Holstein Hamiltonian has both electronic and phonon degrees of freedom. Thus among the questions ML methods can shed light on is which one of these degrees of freedom more clearly encodes the ordered phase.

In this paper, we apply the ``learning by confusion" (LBC) method \cite{van2017learning} to investigate the critical phenomena emerging in the half-filled Holstein model, and map out its phase diagram.
At its heart, LBC consists of a series of supervised learnings with labels that change based on a guess for the location of the critical point as a tuning parameter is varied. 
The correct guess is expected to yield the best accuracy for the training.
LBC has previously been applied to a variety of the fundamental descriptions
of classical magnetic phase transitions, including the
Ising\cite{arnold2023fast,van2017learning}
and XY\cite{lee2019confusion,beach2018machine} models,
as well as the Blume Capel model
where vacancies introduce a first-order line, which is
separated by a tricritical point from the conventional
second order Ising transition \cite{richter2023learning}.
Further applications of LBC include determining the critical value at
which a family of quantum states become
entangled\cite{gavreev2022learning},
phase transitions between distinct steady state behaviors in the dynamics of 
non-linear polariton lattices \cite{zvyagintseva2022machine},
and transitions between regular and chaotic behavior in quantum billiards
\cite{kharkov2020revealing}.
Topological transitions in 
Ising gauge theory and the toric code
\cite{greplova2020unsupervised},
and non-equilibrium quantum quenches captured by experimental images of ultracold atomic gases
described by the one dimensional Bose-Hubbard model \cite{PhysRevLett.127.150504}
are other recent venues where LBC has proven its utility.

Using LBC to explore electron-phonon physics offers the opportunity to 
study issues including (i) whether the fermionic or bosonic snapshots better encode the
CDW phase and (ii) the use of space versus space-time snapshots for the training. We also (iii) use LBC to trace a {\it crossover} from independent gases of up and down spin fermions in the small electron-phonon coupling region
to a spatially random arrangement of empty and doubly occupied sites in the large coupling region
in the absence of CDW order at relatively high temperatures.
This crossover is closely analogous to that which occurs in the Hubbard model as the temperature is lowered and local moments form, but before
those moments order antiferromagnetically.

The remainder of this paper is organized as follows.  In Sec.~\ref{sec:model+methods} we
define the Holstein Hamiltonian and discuss its physics. We also introduce the determinant quantum Monte 
Carlo (DQMC) method, with which we generate snapshots, and the LBC method in some detail.
With this background, in Sec.~\ref{sec:results}, we present results for the CDW transition and local pair crossovers
in the half-filled Holstein model, and discuss
the use of phonon vs.~electron snapshots as well as equal vs.~unequal
time correlators.  Our analysis culminates in a phase diagram of the half-filled
Holstein model in the plane of temperature and the dimensionless electron-phonon coupling. Sec.~\ref{sec:conclusions} presents our concluding remarks.

\section{Model and methods}
\label{sec:model+methods}



\subsection{Holstein Model}
\label{subsec:Holstein}
Interactions between electrons and phonons in materials give rise to dressed quasiparticles (polarons)
of enhanced mass\cite{franchini2021polarons,prokof1998polaron}, as well as collective phenomena such as metal-insulator transitions, superconductivity, 
and charge-ordered states \cite{holstein1959,scalettar89,powell2009introduction,cohenstead20,nosarzewski21, dee2019temperature}.  The Holstein model, given by the Hamiltonian,

\begin{equation}
\hat{H} = \hat{K} + \hat{U} + \hat{V}
\end{equation}
with
\begin{eqnarray} 
\nonumber
\hat{K} &=& -t \sum_{\langle ij\rangle\sigma}
\big(\hat{c}^\dagger_{i\sigma}\hat{c}^{\phantom{\dagger}}_{j\sigma} + h.c.\big)
-\mu \sum_{i\sigma} \hat n_{i\sigma}
\\
\nonumber
\hat{U} &=& \sum_i
\frac{m \omega_0^2}{2} \, \hat{x}_i^2 
+ \frac{1}{2m} \, \hat{p}_i^2
\\
\nonumber
\hat{V} &=& \lambda \sum_{i\sigma} \hat{x}_i \, \big( \hat{n}_{i\sigma} -\frac{1}{2} \big)
\label{eq:hamiltonian}
\end{eqnarray}
is one of the most fundamental tight-binding models for describing electron-phonon interactions. The Hamiltonian comprises 
three components. The first term, $\hat{K}$, represents the nearest-neighbor electron hopping (kinetic energy)
and a chemical potential $\mu$ which controls the electron density. The second term, $\hat{U}$, accounts for the dispersionless
phonon kinetic and potential energy, modeled as a collection of quantum harmonic oscillators. The third term, $\hat{V}$, describes
the on-site electron-phonon interaction, parametrized by the electron-phonon coupling constant $\lambda$.
The electron-phonon interaction term is expressed in a particle-hole symmetric form so that half-filling
$\langle \, \hat n_{i\sigma} \, \rangle = \frac{1}{2}$ occurs at  $\mu=0$,
where also $\langle \hat{x} \rangle = 0$.
We follow the usual convention of setting $m=1$.

In this work we consider a square lattice of linear dimension $L$. The dispersion relation is given by
$\epsilon_\textbf{k}=-2\,t \, (\cos{k_x}+\cos{k_y})$, with a corresponding bandwidth $W=8\,t$. We introduce the dimensionless 
electron-phonon coupling constant $\lambda_D = \lambda^2/(\omega_0^2 \, W)=2g^2/(\omega_0 \, W)$. 
Here $g$ is the coefficient of the electron-phonon coupling when written in terms
of phonon creation and destruction operators
$\hat{V} = g \sum_{i\sigma} (\, \hat{b}^{\phantom{\dagger}}_i + \hat{b}^{\dagger}_{i} \, )\, \big( \hat{n}_{i\sigma} -\frac{1}{2} \big)$.

Ignoring the phonon kinetic energy and then completing the square, one sees that
the Holstein model describes an on-site phonon-mediated attractive interaction between spin up and spin down electrons given 
by $U_{\text{eff}}=-2g^2/\omega_0=-\lambda^2/\omega_0^2 = \lambda_D W$.
This interaction gives 
rise to two notable collective phenomena: (i) A finite-temperature phase transition to charge density wave (CDW) order at
half-filling and (ii) superconductivity upon doping. 
In the former, as temperature decreases, small bipolarons (doubly occupied sites) begin to form. 
Upon reaching the CDW transition temperature, 
these bipolarons, whose number is precisely $L \times L / 2$, 
localize on a preferred sub-lattice, forming a checkerboard pattern.
Two of the foci of this paper are on detecting the CDW phase transition, and showing LBC is also effective at detecting the crossover associated with
the suppression of singly occupied sites (prior to CDW formation which occurs at lower temperature).

\subsection{DQMC Method}
\label{subsec:DQMC}

The snapshots used for training in our LBC method are generated with DQMC~\cite{blankenbecler81,white89}.
In this approach, the partition function ${\cal Z} = {\rm Tr}\,e^{-\beta \hat H}$ 
for the Holstein Hamiltonian is expressed as a path integral
by discretizing the imaginary time $\beta$ into $L_\tau$ intervals and inserting complete sets of eigenstates of the
quantum oscillator position operators.  The trace over the electron degrees of freedom
can be done analytically, and the trace over the phonon degrees of freedom is replaced
by an integral over the oscillator eigenstates $x_{i,\tau}$, which now have both spatial ($i$) and imaginary time ($\tau$)
indices.   The Boltzmann weight has a `bosonic' part, which couples
$x_{i,\tau}$ locally, and so is rapid to evaluate, and a product
of two fermion determinants\cite{bradley21}.  Since the phonon field couples to the two spin species
in the same way, the fermion determinants are identical.  Hence the method has no `sign problem'\cite{loh1990,iglovikov2015geometry,mondaini2022}. 

Within DQMC, there are different methods with which the bosonic fields can be evolved.  In the original
formulation,\cite{blankenbecler81} individual updates at a single space-imaginary time point are proposed.
The locality of this update makes the cost to evaluate the ratio of the new to old Boltzmann weights
scale only as the square of the matrix dimension rather than
its cube, as might naively be expected for a determinant evaluation.  A full sweep of the lattice is then cubic
in system size.  We instead use a variant, `hybrid Monte Carlo', which updates all bosonic degrees of 
freedom simultaneously and is linear in the system size\cite{cohenstead22}.  This method is especially effective for
electron-phonon models, where the phonon kinetic energy moderates the variation of the field in 
imaginary time, and less effective in Hubbard models where such a term is absent\cite{scalettar87hybrid}.

An interesting feature of ML approaches and DQMC for the Holstein model, which we explore below, 
is the possibility of using different types of `snapshots' in the training.  One can, for example, 
present the neural network with the space-imaginary time values of the phonon degrees of freedom
$x_{i,\tau}$.
Alternately, one can utilize the estimator of the density of electrons
$1 - G(i,\tau)$.
Here $G$ is the inverse of the matrix the square of whose determinant is the fermionic
contribution to the Boltzmann weight.
A further flexibility is the ability to restrict to `equal-time' snapshots at a single
$\tau$, rather than snapshots over the entire space-imaginary time lattice.
Finally, one can use correlation functions either of the phonon
degrees of freedom:
$x_{i^{\prime}\,\tau^{\prime}} x_{i,\tau}$,
or of the electrons: $G(i^{\prime},\tau^{\prime}) \, G(i,\tau)$.

\begin{figure}
    \includegraphics[width=1\linewidth]{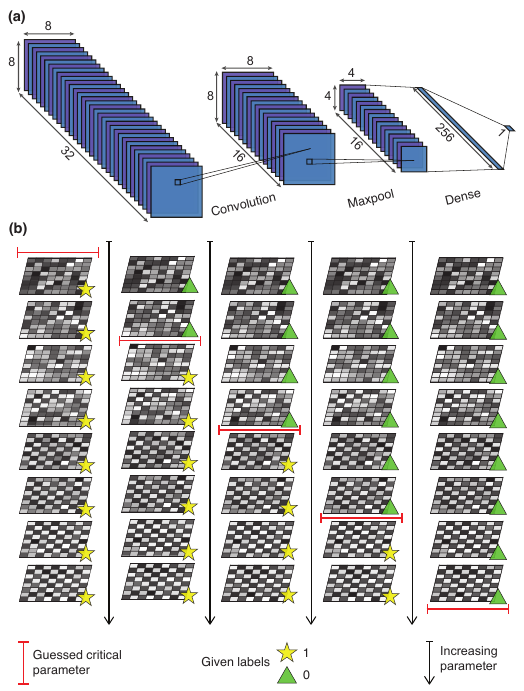}
    \caption{\justifying \textbf{(a)}: Architecture of the CNN used by the LBC technique. The input layer comprises a batch of $32$ samples of size $L \times L$, shown here for $L=8$. There are $16$ kernels (filters) of size $3\times 3$ applied on each data sample in the convolutional layer. The resulting data size is then cut by half after maxpooling. Finally, two dense layers, each with 256 nodes, connect to a single node to output a number between $0$ and $1$ (for simplicity, we show only one of these layers).
    \textbf{(b)}: Schematics of LBC method. 
    Eight different electron density snapshots are shown for increasing values of $\beta$ from the top to the bottom. Here, a white pixel corresponds to $0$ and a black pixel corresponds to $2$. In every step, the LBC 
    algorithm guesses a critical inverse-temperature $\beta_c'$, marked as a red cursor. Data sets above and below 
    $\beta_c'$ are given different labels, marked as a yellow star and a green triangle for 1 and 0, respectively. The first and last steps
     of the algorithm are trivial: all the data sets have the same label and the accuracy to learn is perfect. 
     In the middle column shown, the guessed labels match the actual labels of different phases, leading to
     a high accuracy on the middle peak of the ${\cal W}$ shape of the accuracy vs $\beta$ plot. The second and fourth columns correspond to incorrect
     values, $\beta_c' \neq \beta_c$, and hence lead to low accuracy.}
    \label{fig:CNN_architecture}
\end{figure}

\subsection{Learning by Confusion Method}
\label{subsec:LBC}

In a supervised binary classification problem, each training data sample is paired with a {\it correct} label, $0$ or $1$.
Using a convolutional neural network (CNN), like the one used in our study and presented in Fig.~\ref{fig:CNN_architecture}(a), the 
task is to predict that correct label for as many test samples as possible. The LBC algorithm
involves performing a sequence of supervised binary classification tasks, in which data samples are
provided with modified (possibly {\it incorrect}) labels. 
In this approach, the classification task requires predicting the labels of the test sample,
and the resulting accuracy is used to determine how close the labels are to the
correct ones, and hence the location of the phase transition as a tuning parameter is varied.
This process is illustrated in Fig.~\ref{fig:CNN_architecture}(b).

We illustrate how the LBC technique can be used to find the critical inverse temperature $\beta_c$ in the
square-lattice Holstein model, above which a long-range ordered CDW phase occurs. Our training and test data 
samples are obtained from the hybrid Monte Carlo simulations discussed in Section \ref{subsec:DQMC}. These samples typically 
consist of $L \times L$ 
grids of local observables, such as the electron densities or phonon
positions collected periodically during the measurement step of a simulation. Alternative data sets may
include imaginary time-resolved density-density correlations.

Each DQMC simulation is performed for fixed values of $\beta$, $\lambda_D$, and $\omega_0$, generating snapshots during the course of
$N_\text{meas}=10,000$ sweeps of the space-imaginary time lattice. To determine $\beta_c$, we perform hybrid Monte Carlo simulations on $N_p$ 
different values of $\beta_{\rm min}\le\beta \le \beta_{\rm max}$, keeping $\lambda_D$ and $\omega_0$ constant. Electron densities and phonon position
snapshots are saved every $n_s$ 
measurements, resulting in $N_{\text{snap}}=N_{\text{meas}}/n_{\text{s}}$ configurations for each $\beta$. The resulting data set has dimensions $(N_p\times N_{\text{snap}}, L \times L)$,  
where each
row corresponds to a single snapshot. Density-density correlation datasets are built using snapshots from $n_{\text{it}}$ equally spaced imaginary times and $100$ different Monte Carlo times. 
The resulting dataset has dimension $(N_p\times n_{\text{it}}\times100, L \times L)$. All snapshots are then reshaped into grayscale images and used as input for our CNN.

\begin{figure}
    \includegraphics[width=1\linewidth]{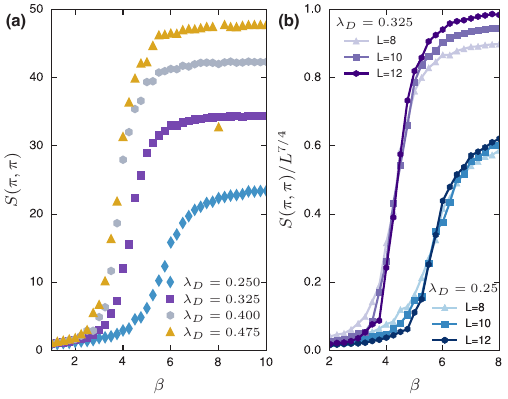}
    \caption{\justifying \textbf{(a)}: The CDW structure factor $S(\pi,\pi)$ for a $\beta$ sweep plotted 
    for four values of electron-phonon coupling, $\lambda_D=0.250, 0.325, 0.400, 0.475$.
    \textbf{(b):} The scaled CDW structure factor plotted for three 
    lattice sizes and the first two values of $\lambda_D$.
     The rapid increase of $S(\pi, \pi)$, and the crossings, happen at values of $\beta_c$ 
     close to the location of the middle peak of the ${\cal W}$ shown in 
    Fig.~\ref{fig:electron_density_acc_lambdaDs}, i.e.~in agreement with the LBC determinations of the critical points.}
    \label{fig:snapshots+DenDen}
\end{figure}

Every grayscale image is labeled based on its corresponding $\beta$. In a typical LBC run, a critical inverse 
temperature $\beta_c^{\,\prime}$ is chosen, and the input labels are modified accordingly: if $\beta \le \beta_c^{\,\prime}$, 
the image is assigned a label of $1$, and if $\beta>\beta_c^{\,\prime}$, it is assigned a label of $0$. A binary 
classification task is then performed using these labels. 
Figure \ref{fig:CNN_architecture}(b) illustrates five key scenarios in the LBC method. 
In the first and last cases, $\beta_c^{\,\prime}=\beta_{\text{min}}$ and $\beta_c^{\,\prime}=\beta_{\text{max}}$, respectively. Here, 
all input labels are identical, and the CNN is trained to label any test set in the same way, a task for which
it easily achieves perfect performance. 

More interesting scenarios arise when $\beta_{\text{min}}<\beta_c^{\,\prime}<\beta_{\text{max}}$. If $\beta_c^{\,\prime}=\beta_c$, 
the CNN performs optimally, as images with fundamentally different data patterns are assigned different labels
{\it which correctly conform to the underlying physics} as contained in the snapshots. 
However, when $\beta_{\text{min}}<\beta_c'<\beta_c$ or $\beta_c<\beta_c^{\,\prime}<\beta_{\text{max}}$, the CNN performs relatively poorly. 
In these cases, the data labels do not align well with the actual patterns. For example, if $\beta_c^{\, \prime} > \beta_c$, 
many high-$\beta$ data sets in the range
$\beta_c < \beta < \beta_c^{\, \prime}$ 
are incorrectly given low-$\beta$ labels. This mismatch leads to the CNN becoming 
`confused', reflected in a failure to distinguish accurately between different data patterns. 

The outcome of the five cases results in the characteristic ${\cal W}$-shaped curve of the LBC method, of which we
will give examples in the next section (Figs.~\ref{fig:L12_accuracies_lambdaD0.25},\ref{fig:electron_density_acc_lambdaDs}).
The `outer' peaks of the ${\cal W}$ originate in the
trivial, uniform label cases, $\beta_c'=\beta_{\text{min}}$ and $\beta_c^{\,\prime}=\beta_{\text{max}}$.
The inner peak of the ${\cal W}$ corresponds to the high accuracy 
of the CNN, which occurs when $\beta_c^{\,\prime}=\beta_c$. This feature enables the LBC method to identify $\beta_c$ within
 a range of $\beta$ values by simply locating the middle peak of the ${\cal W}$. 
The LBC technique  is especially useful in cases where obtaining 
an order parameter is challenging, or when an order parameter does not exist at all,
since only `raw' configurations  of the degrees of freedom are employed.
The LBC technique might also have additional advantages in bypassing the need for finite-size scaling analyses 
of correlation functions (which often have large error bars near the transition) to determine $\beta_c$. 
We will return to this issue in the conclusions.

The relative accuracy with which $\beta_c$ can be determined certainly depends on $N_p$, the number of values between $\beta_{\text{min}}$ and $\beta_{\text{max}}$. We seek a value for $N_p$ that ensures a critical parameter with reasonable accuracy and maintains a sufficient number of samples on both sides of the transition point. We also aim to avoid increasing $N_p$ to the extent that training becomes inefficient. This parameter has been adjusted across different simulations, and we found that a value in the range of $25$-$50$ is adequate for medium-sized intervals like those reported.

\section{Results}
\label{sec:results}

\begin{figure}
    \includegraphics[width=1\linewidth]{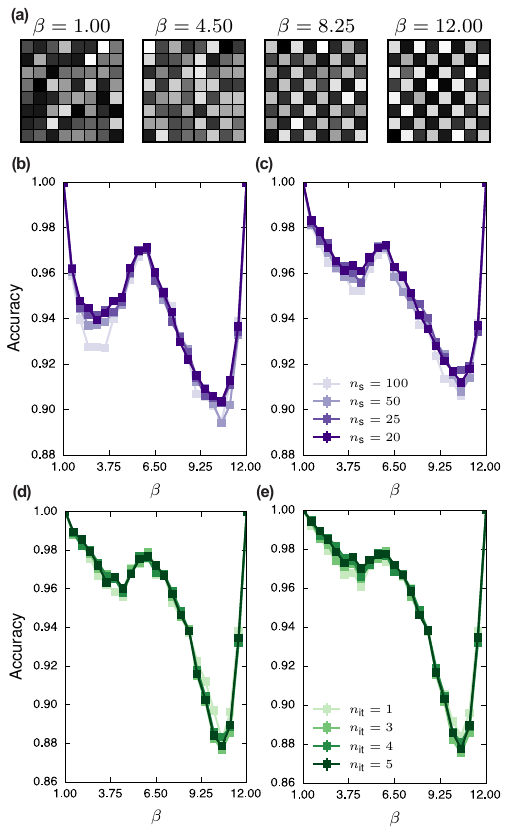}
    \caption{ \justifying Electron density snapshots taken at several temperatures 
    are shown in \textbf{(a)}; the emergence of a CDW pattern is clear at the two larger $\beta$. Also shown are LBC test accuracy for a sweep of temperatures using electron density snapshots \textbf{(b)},
    phonon positions snapshots \textbf{(c)}, density-density correlation data in real space \textbf{(d)}, and in momentum space \textbf{(e)}. 
    The four data types yield consistent positions for the interior maximum of
    the ${\cal W}$ at $\beta_c \sim 5.75$.  Increasing the parameter $n_s$, the number of sweeps between snapshots, reduces the
    number of data sets used in the learning. However, increasing the parameter $n_{\text{it}}$, the number of different imaginary time slices used to build the density-density correlation datasets, increases it. The results for $\beta_c$ are robust to changes in $n_s$, although
    the minima of the ${\cal W}$ deepen slightly as $n_s$ grows.
    Here, $\lambda_D=0.25$, $\omega_0=1.0$, and $L=12$.}
    \label{fig:L12_accuracies_lambdaD0.25}
\end{figure}

It is useful to put the results of LBC in the context of more `traditional' methods for locating the CDW transition.
To this end we show, in  Fig.~\ref{fig:snapshots+DenDen}, the CDW structure factor
\begin{align}
S(\pi,\pi) \equiv \sum_{j} e^{i \pi (j_x+j_y)} \, \langle \, \hat n(j) \hat n(0) \, \rangle,
\end{align}
which sums the density-density correlation at separation $j=(j_x,j_y)$ with a phase
appropriate to ordering of opposite sign on the two sublattices of the bipartite square lattice, vs $\beta$.
At high temperatures, where the correlation function falls off rapidly with separation $j$,
the structure factor is independent of lattice size.  In the ordered phase, correlations extend across
the entire lattice and $S(\pi,\pi) \propto L\times L$.
These two regimes are evident in Fig.~\ref{fig:snapshots+DenDen}(a), with $\beta_c$
roughly estimated as the place where $S(\pi,\pi)$ grows most rapidly.
A more precise determination of $\beta_c$ is obtained by scaling 
$S(\pi,\pi)$ using the known Ising universality class of the CDW transition, for which $\gamma/\nu=7/4$.
Curves for different $L$ cross at $\beta_c$ 
[Fig.~\ref{fig:snapshots+DenDen}(b)].
This is the procedure followed to determine $\beta_c$ in most earlier DQMC
studies of the Holstein model\cite{scalettar89,weber2018,hohenadler19,zhang2019,chen2019,feng2020}.

Fig.~\ref{fig:snapshots+DenDen}(a) shows one under-estimated value for $S(\pi, \pi)$ at $\beta=8.00$ and $\lambda_D=0.475$. This deviation occurs in a challenging strong-interaction and low-temperature regime. However, we note that several data points at even lower temperatures align with the expected trend. We conclude that this deviation does not undermine the validity of the results. We attribute this anomaly to insufficient statistical sampling performed at this point.

\begin{figure}
    \centering
    \includegraphics[width=1\linewidth]{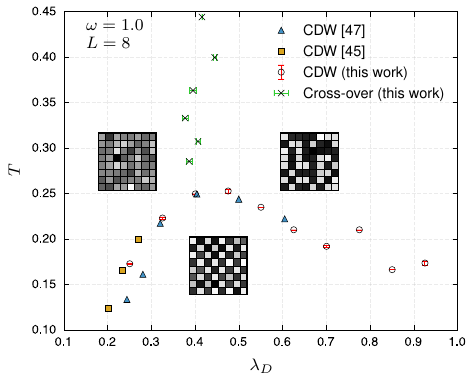}
    \caption{\justifying The square lattice Holstein model $T$-$\lambda_D$ phase diagram obtained with LBC for $L=8$ and $\omega_0=1.0$. The red data points
    indicate the location of $\beta_c$ shown in Fig.~\ref{fig:electron_density_acc_lambdaDs}. Error bars are the standard error of the mean on 10 different random seeds. Critical 
    temperatures of the CDW transition in the range $0.250 \le \lambda_D \lesssim 0.600$ show close agreement with previously obtained results \cite{feng2020, weber2018}. The insets show a typical electron density 
    snapshot taken in each of the three regions of the phase diagram. From left to right, snapshots are taken at $(\lambda_D, \beta)=(0.01, 3.50), (0.25, 12.00), (0.86, 3.50)$. The snapshots show the existence of three distinct phases:
    a disordered phase, a CDW phase, and a Fermi (bipolaron) liquid phase. 
    The anomalous data point at $\lambda_D=0.775$ is discussed in the text.}
    \label{fig:phase_diagram}
\end{figure}

With that standard approach reviewed,
we next present the results using LBC.  We acquire snapshots from a family of simulations at different inverse-temperatures $\beta$ using hybrid Monte Carlo 
simulations
of the Holstein model on a square lattice
 of linear size $L=12$ with $\omega_0=1.0$ and $\lambda=\sqrt{2}$ ($\lambda_D=0.25)$.
 The resulting snapshots are fed into the LBC CNN. We show results of the test accuracy for an inverse-temperature sweep $a(\beta)$, using electron density snapshots, phonon position snapshots, and density-density
correlation data in momentum and position space. Typical electron density snapshots at four different
$\beta$ values are shown in Fig.~\ref{fig:L12_accuracies_lambdaD0.25}(a).
By employing these distinct data types, the LBC results show 
a middle peak of the accuracy $a(\beta)$ for $\beta \sim 6$, agreeing with previously obtained results
\cite{weber2018,hohenadler19}.
The ${\cal W}$ curves obtained from
QMC data for the electron density 
[Fig.~\ref{fig:L12_accuracies_lambdaD0.25}(b)]
and phonon position 
[Fig.~\ref{fig:L12_accuracies_lambdaD0.25}(c)]
snapshots result in deeper
 accuracy minima at either side of the peak, which appear here at $\beta\approx3$ and $\beta\approx10$,
 compared to those observed in the density-density correlations or their Fourier transform, the structure factor
[Figs.~\ref{fig:L12_accuracies_lambdaD0.25}(d), \ref{fig:L12_accuracies_lambdaD0.25}(e)].

\begin{figure*}
    \includegraphics[width=1\linewidth]{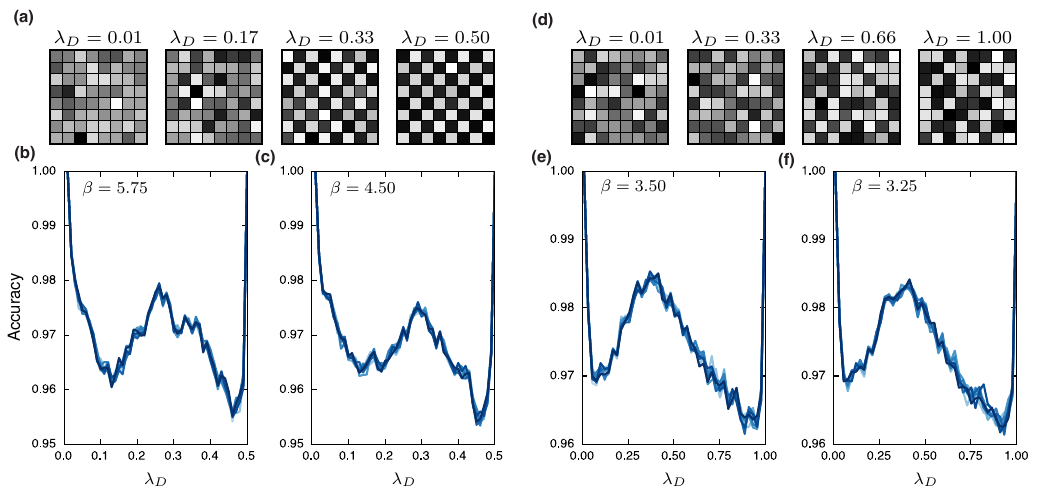}
    \caption{ \justifying 
    Electron density snapshots taken at $\beta=5.75$ are shown in \textbf{(a)}, and the corresponding 
    LBC test accuracy for a sweep of $N_p=50$ values of $\lambda_D$ using them is given in
    \textbf{(b)}.  
    A second sweep of $\lambda_D$ at $\beta=4.50$, also corresponding to a 
   temperature {\it below} the CDW dome, is in \textbf{(c)}. 
   Here, the interior peaks of the ${\cal W}$ mark the onset of long range charge order.
   Phonon position snapshots taken at $\beta=3.50$, shown in \textbf{(d)}, reveal the emergence of long and short bonds without, however,
   an alternating pattern in their positions.
   The corresponding LBC test accuracies for a sweep of $N_p=45$ values of $\lambda_D$ using phonon position snapshots are shown at $\beta =3.50$ \textbf{(e)} and $\beta=3.25$ \textbf{(f)}. Different lines correspond to 10 different random number seeds.
    These temperatures are {\it above} the CDW dome, so the interior maximum of the $\cal{W}$ captures a cross-over
    to a gas of bipolarons. In electron density (phonon position) snapshots, white pixels correspond to $0$ ($-2$) and black pixels correspond to $2$ ($2$). All results are for $\omega_0=1.0$ and $L=8$.} 
    \label{fig:L8_accuracies_cross-over}
\end{figure*}

Figure \ref{fig:L12_accuracies_lambdaD0.25} further explores  the dependence of the LBC results on
the number of training electron density and phonon position data sets; $n_\text{s}=20$ having five times the number of
snapshots as $n_\text{s}=100$.   Although the shape of the ${\cal W}$ away from the interior maximum varies, we observe that the
location of $\beta_c$ is the same. 

Next, we use the electronic density snapshots
to perform similar LBC analyses at other values of $\lambda_D$, ranging 
from 0.250 to 0.925. 
The results are shown in Fig.~\ref{fig:electron_density_acc_lambdaDs}
of the Appendix. The critical 
temperatures we obtain from them paint a complete picture of the CDW phase diagram of the half-filled Holstein model in the space 
of $\lambda_D$ and temperature, which we show
in Fig.~\ref{fig:phase_diagram}. In that figure, circles with red error bars indicate the value of the critical temperature $T_c=1/\beta_c$ for the transition from a disordered 
state at high temperatures, characterized by $S(\pi, \pi) \sim \mathcal{O}(1)$, to the CDW phase at low temperatures, exhibiting 
higher values of $S(\pi, \pi)\sim \mathcal{O}(L^2)$.
We note that the critical temperatures obtained
from the two sweeps in $\beta$ at
$\lambda_D=0.250$ and $\lambda_D=0.325$ agree with
the ones extracted from the finite-size scaling analysis shown in Fig.~\ref{fig:snapshots+DenDen}.
Additionally, Fig. \ref{fig:phase_diagram} demonstrates
the close alignment of the critical temperatures
identified for the CDW transition
in the range $0.250 \le \lambda_D \lesssim 0.600$ with prior studies \cite{weber2018, feng2020}.

There is an anomalous data point in 
Fig.~\ref{fig:phase_diagram} at $\lambda_D=0.775$ for which $T_c$ is evidently over-estimated.  We include it to illustrate limitations
of our current understanding of error estimation in the LBC method.  Critical temperatures extracted from
the structure factor, as in  Fig.~\ref{fig:snapshots+DenDen}, can also sometimes lie substantially off the
expected phase boundary.
However, there are typically indications in the raw data, e.g.~large error bars in the vicinity of the crossing,
which signal the extracted data point might be unreliable.
As can be seen in the $\lambda_D=0.775$ panel of Fig.~\ref{fig:electron_density_acc_lambdaDs}, the `warning sign' appears as a shoulder to the main middle peak in the
shape of the ${\cal W}$, leading to a bump around $\beta=5.8$, presumably marking the actual $T_c\sim0.17$.  
Based on the trends we have seen in the $\lambda_D$ sweeps (see Fig.~~\ref{fig:L8_accuracies_cross-over} below), 
we attribute the unexpected extra peak in this diagram to finite size effects.

It is natural to consider the effectiveness of LBC via a sweep in which $\lambda_D$ is changed at fixed $T$. In a phase diagram of Fig.~\ref{fig:phase_diagram}, this $\lambda_D$ 
sweep is expected to first cross the phase boundary {\it horizontally} into a CDW phase at low enough temperatures. 
The results of the LBC
algorithm using electron density snapshots for $\lambda_D$ sweeps at 
fixed $\beta$ values are presented in 
Fig.~\ref{fig:L8_accuracies_cross-over}. Typical snapshots across different $\lambda_D$'s and at $\beta=5.75$ (below the CDW dome) and $\beta=3.50$ (above the CDW dome) are shown in Fig.~\ref{fig:L8_accuracies_cross-over}(a) and Fig.~\ref{fig:L8_accuracies_cross-over}(d), respectively.  In Fig.~\ref{fig:L8_accuracies_cross-over}(b) for $\beta=5.75$ ($T=0.174$), a
clear ${\cal W}$ shape exhibits a middle peak that is located
at $\lambda_{D_c}\approx 0.259$, in agreement with the results obtained from the $\beta$ sweep performed at a fixed $\lambda_D=0.250$,
revealing a peak 
at $\beta_c\approx 5.75$ 
[see Figs.~\ref{fig:L12_accuracies_lambdaD0.25}(b) and Fig.~\ref{fig:electron_density_acc_lambdaDs}].  
Figure~\ref{fig:L8_accuracies_cross-over}(c) 
shows further consistent results at $\beta=4.50$ ($T=0.222$) for transitioning into the CDW phase at small $\lambda_D$.

The phase diagram of 
Fig.~\ref{fig:phase_diagram} emphasizes that at $T\gtrsim 0.250$, the CDW phase gives way to a disordered phase at 
all values of $\lambda_D$. However, the nature of this disordered phase at small $\lambda_D$ ($\lesssim 0.4$) is very 
different from that at large $\lambda_D$.
In the former region,
there is mixture of empty, singly-occupied, and doubly-occupied sites, in which the entropy per site
achieves its maximal value, ${\rm ln}\,4$.
However, for large values of $\lambda_D$ ($\gtrsim 0.5$), bipolarons are preferentially formed, and a gas of mostly empty and 
doubly-occupied sites exists across the lattice, a regime where the entropy per site is ${\rm ln}\,2$.
Typical electron density snapshots shown at a relatively high temperature of $T=0.286$ in Fig.~\ref{fig:phase_diagram} clearly display these behaviors.
Their signature is also reflected in the value of the CDW structure factor in those regions. The latter is shown as a function of $\lambda_D$ at $\beta=5.75$ in Fig.~\ref{fig:DenDen_lambdaD} of the Appendix. 
While $S(\pi,\pi)$ is minimal around $1$ in the completely disordered region of small $\lambda_D$, it saturates to a value around three times as much in the large-$\lambda_D$ region.

The phonon snapshots of Fig.~\ref{fig:L8_accuracies_cross-over}(d)
show how the electron-phonon bond strengths evolve in this cross-over. Note that the right-most panel
of Fig.~\ref{fig:L8_accuracies_cross-over}(d) [at $\lambda_D=1.00$ and $\beta=3.50$ $(T=0.286)$]
correspdonds to empty and doubly occupied sites without the CDW pattern 
of the ordered phase [e.g., the latter is shown in the
right-most panel 
in Fig.~\ref{fig:L12_accuracies_lambdaD0.25}(a) at $\lambda_D=0.25$ and $\beta=12.00$ $(T\approx0.083)$].
It is the {\it cross-over} between the completely disordered and bipolaron regimes that is captured by our LBC analysis of
$\lambda_D$ sweeps at moderate 
temperatures above the CDW dome. Figure~\ref{fig:L8_accuracies_cross-over}(e) and 
~\ref{fig:L8_accuracies_cross-over}(f) show such sweeps for $\beta=3.50$ ($T=0.286$) and $\beta=3.25$ ($T=0.308$), 
respectively. We use phonon field snapshots for the LBC algorithm to obtain these results. 
The corresponding cross-over temperatures are added to the phase diagram of Fig.~\ref{fig:phase_diagram} 
as crosses with green error bars. We find that using electronic 
snapshots in this case
leads to weaker results (see Fig. \ref{fig:density_acc_crossover}). 
Instead of a clear middle peak in the accuracy vs $\lambda_D$
plots, we observe a shallow minimum in the small-$\lambda_D$ region
and a broad middle peak, followed by a relatively sharp minimum 
in the large-$\lambda_D$ region.
 
The ${\cal W}$ curves for the $\lambda$ sweeps 
of Fig.~\ref{fig:L8_accuracies_cross-over}
are notably more noisy than the
$\beta$ sweeps of 
Figs.~\ref{fig:L12_accuracies_lambdaD0.25} and
\ref{fig:electron_density_acc_lambdaDs}.
We attribute this to our finite lattice size, and hence, the coarse resolution of the
Brillouin zone.  On a finite lattice and at low temperatures, the electron density exhibits 
a step-like structure as the chemical potential passes through the energies of the discrete $k$ points.
These fictitious jumps are removed by sufficiently strong interactions.  However, 
our horizontal sweeps begin at rather small $\lambda$, where finite size effects are large.
We believe these then get reflected in the appearance of subsidiary structure in the ${\cal W}$.
The smoother ${\cal W}$ curves of 
Figs.~\ref{fig:L12_accuracies_lambdaD0.25} and
\ref{fig:electron_density_acc_lambdaDs}
all use larger $\lambda_D$, suppressing finite size effects.

The crossover which is captured by the LBC method in Fig.~\ref{fig:phase_diagram}
can be understood as follows:
The half-filled repulsive Hubbard model is well-known to exhibit 
two peaks in its specific heat as the temperature is lowered\cite{paiva01signatures}.
These correspond to, at higher temperature, the formation of local moments (singly occupied sites)
and the entropy loss as empty and doubly occupied sites are removed from the system,
and at lower temperature to antiferromagnetic ordering of those moments.
Indeed, a phase diagram similar to our Fig.~\ref{fig:phase_diagram}
then results \cite{chng17machine}.
As a consequence of a well-known particle-hole transformation\cite{singh91exact}, the specific
heat of the attractive Hubbard model has a similar structure, reflecting first the elimination of
single occupied sites and then CDW formation as $T$ decreases.
It is also known that the Holstein model maps onto the attractive Hubbard model
in the anti-adiabatic limit $\omega_0 \rightarrow \infty$.
In that sense, the phase diagram obtained from LBC is a plausible one.
However, it has been shown that to achieve the limit in which Holstein quantitatively
maps onto attractive Hubbard, $\omega_0/t \sim 100$ is required
\cite{feng20charge}.
This is far from the $\omega_0/t=1$ studied here.
Thus, the observation of a crossover to a regime of randomly arranged empty and doubly occupied sites
is a notable achievement of the LBC approach.

The LBC method also proves highly effective in the adiabatic regime of the Holstein model, $\omega_0 \to 0$. In this limit, determining $\beta_c$ is particularly challenging due to pronounced finite-size effects. Traditionally, finite-size scaling must be performed on relatively large lattice sizes to extract $\beta_c$ with reasonable accuracy. To illustrate the usefulness of LBC in this limit, we perform a $\beta$ sweep at fixed $\lambda_D=0.250$ and $\omega=0.1$. The accuracies, obtained from electron density snapshots, are shown in Fig.~\ref{fig:density_acc_omega0.1}.

\bigskip
\section{Discussion} 
\label{sec:conclusions}

In this paper we have studied the CDW transition of the Holstein model, and the crossover to
a gas of small polarons, with learning
by confusion.  One focus of our investigation was on the relative effectiveness of using
spatial snapshots  at a single imaginary time slice, versus
using the full space-imaginary time lattice.
In the case of the transverse field Ising model in $d$ dimensions,
the path integral mapping of the partition function is to a classical Ising
model in $d+1$ dimensions.  In that case, withholding the imaginary time
direction is precisely a matter of using snapshots only on a $d$ dimensional hyperplane
embedded in a larger $d+1$ dimensional lattice.  However, the simplicity of the transverse field
Ising model mapping is atypical.  In general, and here in the Holstein model in particular,
the space and imaginary time directions behave very differently.  Hence, our results speak to
the complex, and more generic situations when  this is the case.

A commonly enunciated advantage of the LBC method is its ability to work directly
with snapshots as opposed to requiring a structure factor or a particular (possibly unknown)
order parameter.  However, this is a property shared by a number of ML approaches\cite{johnston2022perspective}.
Indeed, principal component analysis also works directly with snapshots, with the additional feature that the
leading eigenvector of the covariance matrix of data returns information about the order parameter\cite{hu2017discovering}.
While we have not done a careful study,
our results suggest another possible advantage, namely relatively small finite size effects. 
Further work is needed to understand how the position of the inner peak of the ${\cal W}$,
which encodes the critical point, depends on the lattice size, in analogy with techniques which
have developed over the last few decades for the finite size scaling
of peaks in the specific heat and susceptibilities, as well as invariant quantities
such as Binder ratios\cite{binder1981finite,binder1984finite,sandvik1997finite}.
In the course of such a comparison, the question of whether LBC allows the determination
of critical points with higher accuracy than `traditional' approaches (allowing for the analysis of both
statistical and systematic error bars) can be better understood.  
 Our current results suggest that LBC
is a more efficient approach (using less CPU time) to obtaining the phase diagram of the 
Holstein Hamiltonian to the presented level of precision (a few percent uncertainty in $\beta_c$). 

In this text, we have portrayed the ability of LBC to detect both a real phase 
transition and a crossover. However LBC does not retain specific underlying 
structures of the data that could be directly interpreted physically. Instead, at 
its highest level, LBC simply maps an input configuration to a label, namely $0$ or
$1$. Therefore, the current implementations of LBC do not distinguish between a real
phase transition and a crossover in the same rigorous manner as the sophisticated 
finite-size scaling methods that have been developed. However, one viable approach
to distinguish between a phase transition and a crossover with LBC
is to examine how the 
critical temperature $T_c$, identified by the position of the interior peak in the 
${\cal W}$, depends on 
linear lattice size: the functional form of the
correction $T_c(L) = T_c(\infty) + A L^{-1/\nu}$ might then allow access to the 
critical exponent 
$\nu$ and the nature of the transition.

While density-density correlation data might be expected to yield the most accurate $\cal W$, our findings demonstrate that raw QMC data, 
such as electron densities and phonon positions, generally produce a clearer $\cal W$ with deeper minima. 
The choice between using electron densities or phonon positions depends on the specific parameters of the problem. 
For instance, electron density snapshots yielded a clearer $\cal W$ when sweeping $\beta$ to explore the CDW transition, 
whereas phonon position snapshots proved more effective for analyzing the crossover during a $\lambda_D$ sweep. 
The ability of the LBC method and other ML approaches to only require snapshots already eliminates the need of considering 
to feed correlation snapshots into the model. This is promising since to an extent, extracting the location of a phase transition 
from correlation data is not too far from giving the machine the answer we expect from it. 

We conclude by noting a subtle feature of the LBC method.
It is evident in the results presented in this work that
the minima in  the ${\cal W}$ curve can typically be not so much reduced from unity; in several of our plots
the {\it lowest} accuracy is as high as 0.93, despite the fact that those $\beta$ values
correspond to the data sets being the `most mislabeled', i.e.~$\beta_c'$ very far from
the correct $\beta_c$. 
The explanation reflects the power of ML and training.  Given enough time (epochs) and fitting parameters (weights and biases),
a CNN should ultimately be able to learn to classify test sets according even to `incorrect' labels. 
Thus in some sense the minima in the ${\cal W}$ curves are 
reliant on the limitation of resources. \cite{lee2019confusion}

\begin{acknowledgments}
This work was supported by the 
grant DE‐SC0022311 funded by
the U.S. Department of Energy, Office of Science.
\end{acknowledgments}

\section*{Data Availability Statement}
The data that support the findings of this article are openly available.\footnote{Available in our \href{https://ucdavis.app.box.com/folder/304696521244}{data repository}.}

\bigskip

\newpage
\renewcommand{\thefigure}{A\arabic{figure}}
\setcounter{figure}{0}

\section{Appendix}

\begin{figure*}
    \centering
    \includegraphics[width=1\linewidth]{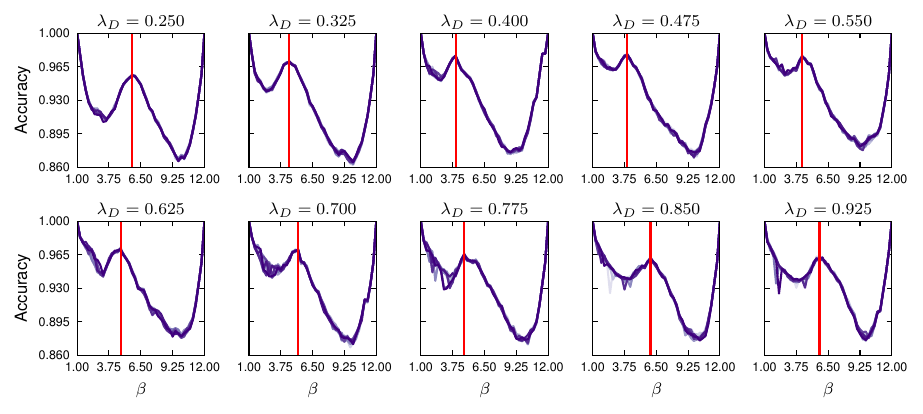}
    \caption{\justifying The test accuracy of the LBC model for a sweep of $\beta$ using electron density 
    snapshots obtained for 10 different values of $\lambda_D$. We take $\omega=1.0$ and $n_{\text{s}}=20$, and $N_p=45$.
     The red vertical line is drawn at the location of the average middle peak of the ${\cal W}$, taken over the $10$ random number seeds shown.}
    \label{fig:electron_density_acc_lambdaDs}
\end{figure*}

\begin{figure}
    \centering
    \includegraphics[width=1\linewidth]{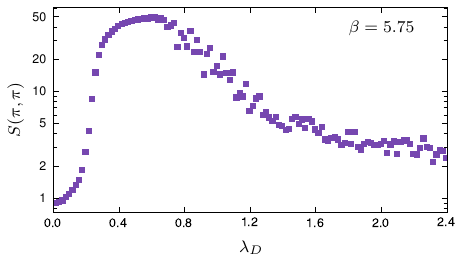}
    \caption{\justifying The structure factor against $\lambda_D$ at $\beta=5.75$ plotted on a semi-log scale. $S(\pi, \pi)$ shows a visible difference between the completely disordered phase at low $\lambda_D$ and the bipolaron structure at high $\lambda_D$.}
    \label{fig:DenDen_lambdaD}
\end{figure}

\begin{figure}
    \centering
    \includegraphics[width=1\linewidth]{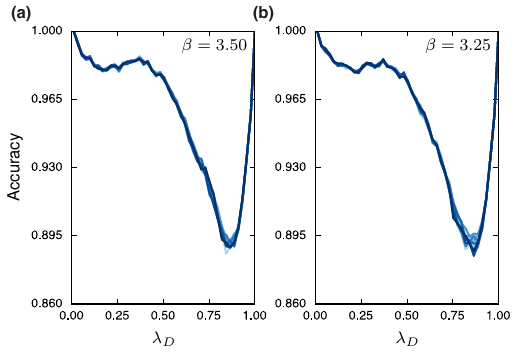}
    \caption{\justifying The LBC test accuracy for a sweep of $\lambda_D$ at $\beta=3.50$ \textbf{(a)} and $\beta=3.25$ \textbf{(b)} using electron density snapshots. Both temperatures lie {\it below} the CDW dome. Different lines correspond to 10 different random
    number seeds.}
    \label{fig:density_acc_crossover}
\end{figure}

\begin{figure}
    \centering
    \includegraphics[width=1\linewidth]{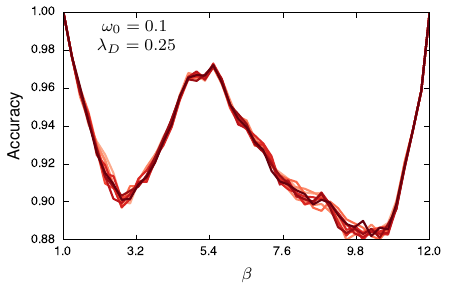}
    \caption{\justifying The test accuracy of the LBC model for a sweep of $\beta$ using electron density snapshots. Here, $\omega=0.1$, $n_s=100$, and $N_p=45$. The success of LBC at this low value of $\omega_0$ portrays the usefulness of LBC in the adiabatic limit, $\omega_0 \to 0$.}
    \label{fig:density_acc_omega0.1}
\end{figure}

We repeat the same $\beta$ sweep
in the main text for $10$ different values of $\lambda_D$ and show the results in Fig.~\ref{fig:electron_density_acc_lambdaDs}. 
Here we use a lattice size $L=8$.
Indeed, we see very little variation of the ${\cal W}$ curves with lattice size 
(compare, for example, Fig.~\ref{fig:L12_accuracies_lambdaD0.25}
with the top left panel of 
Fig.~\ref{fig:electron_density_acc_lambdaDs}, which shows results for the same
$\lambda_D=0.25$.) 
The red 
vertical lines in Fig.~\ref{fig:electron_density_acc_lambdaDs} represent the location of $\beta_c$ for the 10 different values of $\lambda_D$. 

In Fig.~\ref{fig:DenDen_lambdaD}, we show the CDW structure factor as a
function of $\lambda_D$ at $\beta= 5.75$ in a semi-log plot. It shows that $S(\pi, \pi)$ is larger, by a about a 
factor of three, in the large-$\lambda_D$ region, where the system is expected to consist of mostly a gas of empty
and doubly-occupied sites, in comparison to the completely disordered small-$\lambda_D$ region before the peak.

In Fig.~\ref{fig:density_acc_crossover}, we show the same plots as in Fig.~\ref{fig:L8_accuracies_cross-over}(e) and \ref{fig:L8_accuracies_cross-over}(f) of the main text, except that electronic snapshots, as opposed to phonon position snapshots, are used. In this case, we find a broader peak and a shallower minimum in the small-$\lambda_D$ region.

\end{document}